# Deep Learning Classification of EEG Responses to Multi-Dimensional Transcranial Electrical Stimulation

Alexis Pomares Pastor[1,*], Ines Ribeiro Violante[1,2], Gregory Scott[1,3]

[1] Centre for Neurotechnology, Imperial College London

[2] School of Biomedical Engineering & Imaging Sciences, King's College London

[3] Neurology Department, UK Dementia Research Institute

[*] *Corresponding author:* alexispomarespastor@gmail.com

## ABSTRACT

A major shortcoming of medical practice is the lack of an objective measure of conscious level. Impairment of consciousness is common, e.g. following brain injury and seizures, which can also interfere with sensory processing and volitional responses. This is also an important pitfall in neurophysiological methods that infer awareness via command following, e.g. using functional MRI or electroencephalography (EEG).

Transcranial electrical stimulation (TES) can be employed to non-invasively stimulate the brain, bypassing sensory inputs, and has already showed promising results in providing reliable indicators of brain state. However, current non-invasive solutions have been limited to transcranial magnetic stimulation, which is not easily translatable to clinical settings. Our long-term vision is to develop an objective measure of brain state that can be used at the bedside, without requiring patients to understand commands or initiate motor responses.

In this study, we demonstrated the feasibility of a framework using Deep Learning (DL) algorithms to classify EEG brain responses evoked by a defined multi-dimensional pattern of TES. We collected EEG-TES data from 11 participants and found that delivering transcranial direct current stimulation (tDCS) to posterior cortical areas targeting the angular gyrus elicited an exceptionally reliable brain response. For this paradigm, our best Convolutional Neural Network models reached a 92% classification F1-score on Holdout data from participants never seen during training, significantly surpassing human-level performance at 60-70% accuracy.

We hope that our findings will pave the way for further experiments with healthy participants in asleep states as well as with patients in abnormal states of consciousness, and that our insights into TES-evoked brain function will contribute to developing a robust measure of conscious level that could transform clinical decision making in patients with disorders of consciousness. In this spirit, we documented and open-sourced the project in full—including datasets and commented code—to be used freely by the neuroscience and AI research communities, who may replicate our results with free tools like GitHub, Kaggle, and Colaboratory.

| 92% | 60-70% | 11 |
|---|---|---|
| ACCURACY (HOLDOUT DATA) | HUMAN-LEVEL ACCURACY | PARTICIPANTS |

## 1. Introduction

Brain imaging technologies are a common tool in neuroscience research environments, but they present inherent limiting factors that have prevented wide adoption in clinical settings. Electroencephalography (EEG) provides a good balance between cost-effectiveness, reliability, convenience, and signal quality. Nevertheless, EEG yields a complex signal that can require years of training to master its subjective interpretation, and advanced signal processing and feature extraction techniques to implement into functional applications[1]. Clinical practice today relies on one technical specialist to record EEG, and a separate neurophysiologist to analyse it and draw conclusions on patient's brain state, a process that can take as long as several weeks[2].

The overarching long-term goal of our research is to develop a novel measure of brain state that can be easily used at the patient's bedside, derived from the set of brain responses evoked by a multi-dimensional (potentially varying in intensity, time and space) pattern of electrical neurostimulation. We want to leverage Artificial Intelligence (AI) to reliably automate —expedite and facilitate—the assessment of consciousness level, removing exposure to human error and simplifying the workflow of clinicians.

### 1.1 Measuring Consciousness

The lack of a 'gold standard' to measure conscious level is a major shortcoming in medical practice[3,4]. Bedside evaluation of consciousness relies on a patient's ability to hear or see, understand, and act on instructions. Unfortunately, impairment of consciousness is common following brain injury or seizures[5], and can often be accompanied by impaired sensory processing and volitional responses, a major confound for the behavioural assessment of conscious level[6]. Current approaches can therefore miss clinically significant abnormalities in patient's states of consciousness—an important concern that could be solved with an objective measure that can bypass processing of verbal commands and motoric responses.

### 1.2 Functional Brain Imaging

Researchers sought such a measure by turning to non-invasive functional brain imaging methods, such as functional magnetic resonance imaging (fMRI) and EEG[7]. While fMRI achieves high signal-to-noise ratio and spatial resolution (typically 3–4 mm), its temporal resolution is comparatively poor (around 1–2 seconds) due to hemodynamic response limitations[8]. More importantly, fMRI lacks clinical scalability due to relatively expensive equipment, specialized personnel requirements, and restrictive patient compatibility limitations.

By contrast, EEG measures voltage fluctuations in the scalp with excellent temporal resolution in the order of milliseconds, given it is a direct electrical recording[9]. EEG is widely available, inexpensive, easy to administer at the bedside, fairly robust to many artifacts, and has virtually no restrictions with regard to patient compatibility and safety[7].

### 1.3 Non-Invasive Brain Stimulation

Neurostimulation is a valuable tool to use in combination with brain imaging methods, providing a framework to consistently elicit responses in the brain[4]. Transcranial Magnetic Stimulation (TMS) uses electromagnetic induction to non-invasively perturb cortical activity, and has been successfully used with EEG to produce reliable indicators of conscious level[10]. However, TMS equipment is costly, poorly portable, and complex to setup.

Transcranial electrical stimulation (TES) encompasses techniques such as direct current (tDCS), alternating current (tACS) or random noise (tRNS) used to safely pass electrical current through the cortex[11]. TES is a portable, cost-effective, non-invasive method of directly stimulating the brain, making it a promising alternative to probe brain function in patients with Disorders of Consciousness (DoC) who are not responsive to external stimuli[6].

### 1.4 Neural Correlates of Consciousness

A significant amount of research has converged to identify a "posterior cortical hot zone" as a reliable full Neural Correlate of Consciousness (NCC), right at the intersection between the parietal, occipital and temporal lobes[3,4,12,13]. Recently it has been shown that conscious level is intrinsically linked to the complexity of brain activity, and that it can be detected at rest[14]. States of reduced conscious level are associated with reduced brain complexity, decreased variability in the repertoire of cortical activity, and a reduction in information transfer between cortical regions.

### 1.5 Deep Learning and EEG

In the past decade there has been an increasing trend to explore Deep Learning (DL) techniques to analyse neurophysiological signals, most commonly EEG for its convenience and availability[1,15]. In a systematic review of more than 150 papers, Roy et al.[1] found that different DL techniques applied directly to raw EEG time series accomplished a median gain in accuracy of 5.4% compared to more traditional ML methods. Lee and colleagues[16] used a Convolutional Neural Network (CNN) to achieve 90.9% accurate measure of sleep consciousness, while Ullah et al.[17] developed a DL-based system to detect epileptic seizures with 99.1% minimum accuracy.

### 1.6 Research Objective

For this pilot study, our aim was to answer the following research question: *Can a Deep Learning algorithm distinguish the EEG brain responses evoked by different types of electrical stimulation in distant cortical regions?*

Based on controllability analysis from Deco et al.[18], we targeted two bilateral neuroanatomic regions of interest (ROIs): the angular gyrus (posterior areas) with highest controllability, and the middle frontal gyrus (anterior regions) with low controllability. We delivered TES in two modalities—tDCS and tACS—across these spatial conditions.

## 2. Methods

### 2.1 Equipment

We used a GTEN 200 neuromodulation system (Electrical Geodesics Inc, EGI) that allows TES—in any combination of tDCS/tACS/tRNS—and high-density EEG to be recorded through the same 256-electrodes cap[19]. The GTEN allows stimulation through any combination of electrodes whilst simultaneously recording EEG from remaining electrodes.

### 2.2 Experimental Design

Each session was split into 10 runs with 14 individual blocks, alternating stimulation periods of 2mA peak intensity with non-stimulation periods. Rest blocks are intervals when EEG was recorded without stimulation, classified as either 'M' (Measure—blocks immediately after stimulation) or 'C' (Control—rest periods following 'M' blocks when TES effects were expected to be minimal).

We targeted cortical regions by designating a narrow 5-channel circular anodal ring directly above each ROI, with a larger concentric cathodal ring surrounding the anodes with a separation ≥3 centimetres. This montage ensured maximal TES current would flow focally through our brain ROIs.

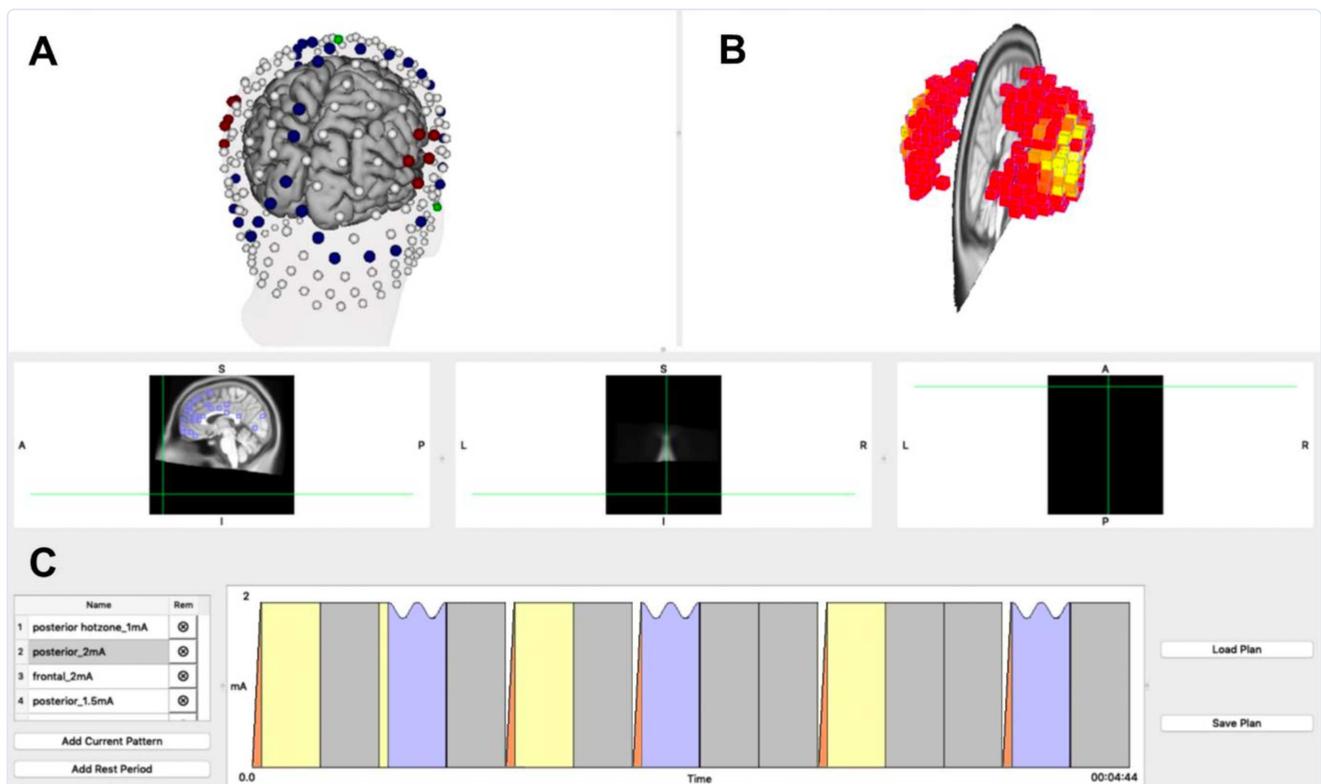

**Figure 1.** GTEN planning software interface. (A) Designation of electrodes to deliver anodal stimulation (red) and corresponding cathodes (blue). Note the bilateral concentric ring shape to maximize focal current intensity. (B) Estimated current intensity for EGI's brain atlas, showing maximum intensity (yellow) in narrow cortical areas beneath anode electrodes. (C) Example of a non-calibration run showing tDCS (yellow), Rest (grey) and tACS (purple) blocks.

## 2.3 Participants and Data Collection

We enrolled 11 healthy resting awake participants (4 female; ages 20-37, average 25.0±4.6 years) to conduct 13 separate experimental sessions. Participants were instructed to sit awake with eyes open, and blinded to the conditions applied. Following an initial rest period of 120 seconds, up to 58 blocks of TES were performed, with a total time of up to ~60 minutes per session. EEG was continuously recorded at a sampling rate of 1000Hz.

All participants gave written informed consent. The study conforms to the Declaration of Helsinki, and ethical approval was granted through the local ethics board (IRAS Project ID: 154511).

## 2.4 EEG Preprocessing

We developed a holistic pipeline in Python using the MNE library to preprocess, visualize and analyse raw EEG data[20,21]. First, we removed the reference electrode and 68 'bad' channels that consistently had poor contact with the scalp. We resampled signals from 1000Hz to 250Hz to expedite training.

Next, we filtered the downsampled data (1Hz high-pass, 80Hz low-pass, 50Hz notch) to reduce common EEG artifacts such as DC baseline drift, high-frequency noise, and electrical grid contamination. We then epoched the continuous EEG time series into non-overlapping windows of 1000 milliseconds.

## 2.5 Feature Extraction

We computed 37 electrode-wise statistical features along the time axis for each epoch, including: basic statistics (mean, median, std, max, min), standard EEG power measures across 5 bands (delta, theta, alpha, beta, gamma) in both linear and logarithmic scales[22], and 22 features from the catch22 package[23]—high-performance measures selected for their ability to capture structural information of time series data.

We produced two datasets: timeseries (250 samples × 188 electrodes) and features (37 statistics × 188 electrodes).

## 2.6 Deep Learning Analysis

We partitioned data into Training, Validation and Holdout Test datasets. Training and Validation sets were generated from aggregated data from 10 participants (shuffled, randomly 95:5% split). Data from the remaining participant was reserved for Holdout testing (leave-one-out cross validation).

Multiple DL models were developed using TensorFlow[24]. We trained these algorithms to classify EEG responses (Rest blocks) according to their evoking pattern of stimulation (preceding Stim blocks). Starting with CNN architectures inspired by Ashford et al.[25], we explored various topologies including GRU, LSTM, and VGG-16 adaptations.

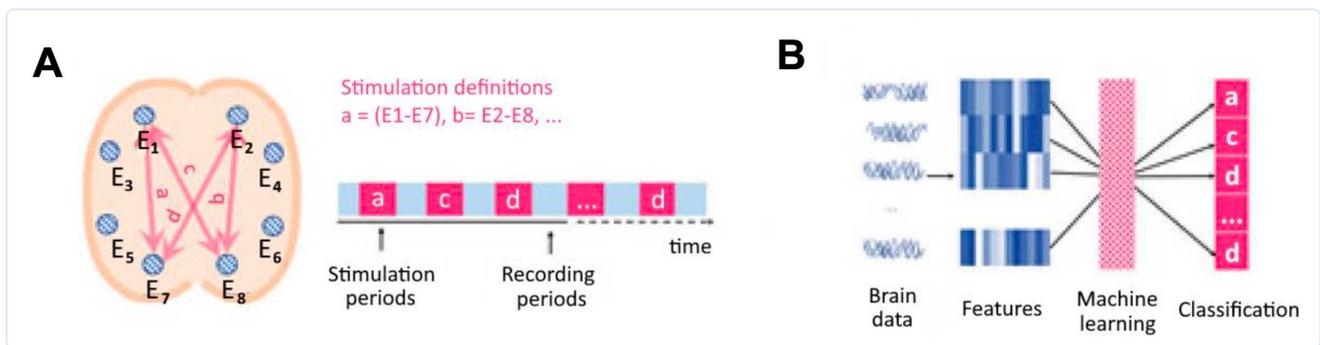

**Figure 2.** (A) GTEN system executes a preset program of brain stimulations—covering a spatially-distributed set of electrode configurations—and records EEG responses following each stimulation. (B) Collected brain data is analysed and fed to a DL algorithm that classifies each signal according to its specific evoking TES pattern.

# 3. Results

### 3.1 Participant Reports and Data Quality

Participants reported mild to moderate sensations of itching, skin redness, visuals, and fatigue/sleepiness. Visual experiences were always phosphenes—participants perceived quickly flickering lights (at 10Hz) caused by tACS current leaking to the retina[26]. During calibration runs, all subjects reported no serious discomfort for 2mA intensity except participant P004, who received a peak amplitude of 1mA.

Analysis of auditory oddball exercises showed reaction times remained stable within 0.37±0.12 seconds, indicating that experimental conditions did not cause significant fatigue or loss of attention.

### 3.2 EEG Preprocessing and Dataset Generation

Our preprocessing pipeline successfully cleaned EEG signals, as shown by Power Spectral Density analysis. After filtering, power density was evenly distributed across frequencies, with grid noise at 50Hz nearly eliminated.

Due to TES-related electrical contamination (mV intensity corrupting μV-scale EEG), we discarded all EEG epochs from time intervals impacted by neurostimulation artifacts—resulting in approximately 52% data loss. This produced a grand total of 166 minutes (2.8 hours) of valid EEG training data, yielding nearly 10,000 training examples. The timeseries dataset contained 2.5 million rows (~3.69GB), while the features dataset consisted of 0.36 million rows (~0.56GB).

### 3.3 Deep Learning Classification Performance

From different DL architectures explored, we regularly observed poor performance with RNN topologies and VGG-16 adaptations, unable to produce accuracies above the ~28% chance level for our imbalanced 5-label classification task.

CNN architectures based on the systematic review by Roy et al.[1] showed great promise, quickly reaching 50% accuracy with minimal modifications. After extensive hyperparameter tuning, we achieved >90% accuracy on Training and Validation datasets.

```
Model: "timeseries-CNN"
_________________________________________________________________
Layer (type)                 Output Shape              Param #
=================================================================
conv2d (Conv2D)              (None, 251, 188, 32)      320
_________________________________________________________________
conv2d_1 (Conv2D)            (None, 251, 188, 64)      18496
_________________________________________________________________
max_pooling2d (MaxPooling2D) (None, 125, 94, 64)       0
_________________________________________________________________
flatten (Flatten)            (None, 752000)            0
_________________________________________________________________
dense (Dense)                (None, 64)                48128064
_________________________________________________________________
dense_1 (Dense)              (None, 5)                 325
=================================================================
Total params: 48,147,205
Trainable params: 48,147,205
Non-trainable params: 0
```

```
Model: "features-CNN"
_________________________________________________________________
Layer (type)                 Output Shape              Param #
=================================================================
conv2d_26 (Conv2D)           (None, 37, 188, 32)       320
_________________________________________________________________
conv2d_27 (Conv2D)           (None, 37, 188, 64)       18496
_________________________________________________________________
flatten_13 (Flatten)         (None, 445184)            0
_________________________________________________________________
dense_26 (Dense)             (None, 256)               113967360
_________________________________________________________________
dense_27 (Dense)             (None, 5)                 1285
=================================================================
Total params: 113,987,461
Trainable params: 113,987,461
Non-trainable params: 0
```

**Table 1.** Network architectures and parameters for timeseries and features classification tasks.

Analysis of confusion matrices revealed that the DL algorithm classified tDCS conditions almost perfectly (98% F1-score), while performance dropped moderately for resting state (87%) and tACS conditions (91%). This suggests that EEG responses to tACS stimulation may resemble resting state neural dynamics at some level.

### 3.4 Holdout Performance and Generalization

Initial Holdout testing resulted in poor performance (~40% accuracy), not significantly above the 28% chance level. DL algorithms excel at learning accurate statistical representations from seen distributions but struggle with out-of-distribution data[27].

We studied regularization techniques including Dropout and L2 Regularization, but found that simply reducing CNN architecture size produced the best results. With smaller models (~1 million vs. 300 million trainable parameters), overfitting was displaced to the 20th epoch and beyond, while Training/Validation accuracies remained >85% and Holdout values crossed the 60% mark.

**Best performance was achieved with the statistical features dataset, reaching a Holdout accuracy of 68.1%** on a CNN model trained for 17 epochs. For timeseries classification, the highest accuracy was 60.5% after 23 epochs.

### 3.5 Neural Network Interpretability

We analysed brain responses across three dimensions: time-frequency analysis, feature importance, and spatial distribution. Spectrograms revealed that resting state showed a relatively flat distribution of frequency power, similar to tACS conditions. In contrast, tDCS conditions showed well-defined spikes of frequency power, appearing almost exclusively in the first half (≤8 seconds) of time intervals.

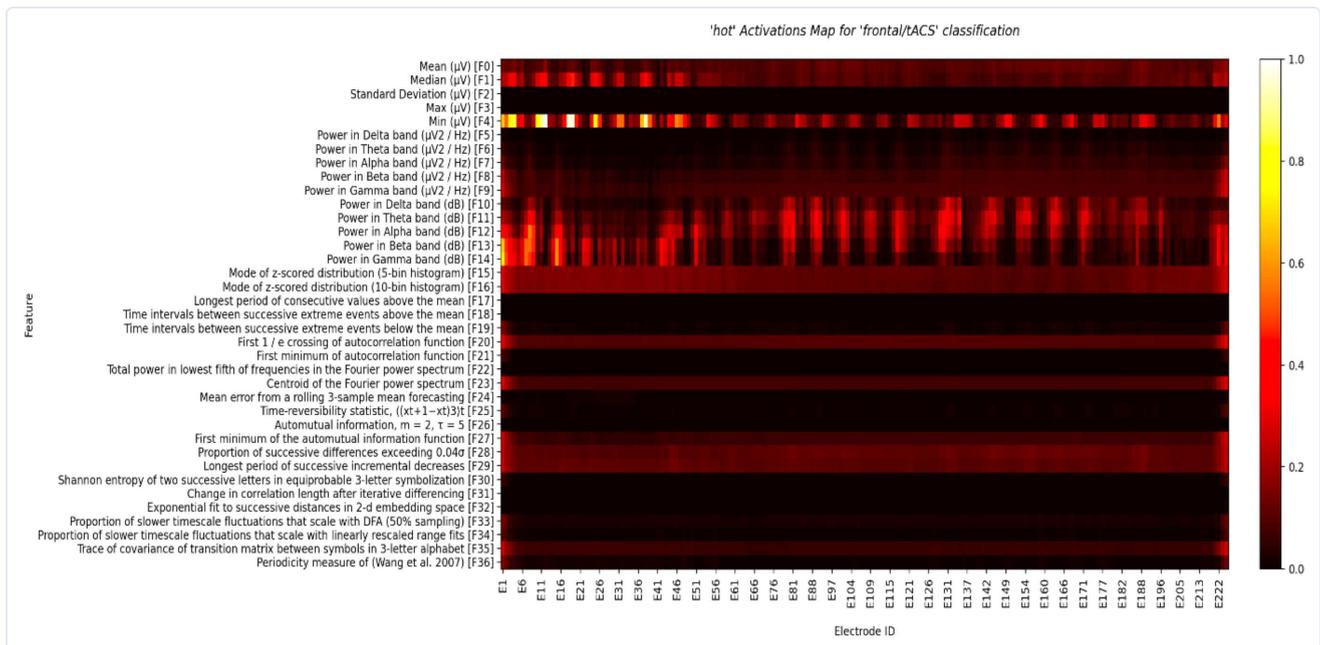

**Figure 3.** Activation map produced with the Grad-CAM technique[28], showing normalized average activations for Holdout data. Max-intensity pixels correspond to feature+electrode combinations with largest impact on predictions.

Using Gradient-weighted Class Activation Mapping (Grad-CAM)[28], we found the most important features were common statistics (median, minimum, mean), mode of z-scored distributions, and power in alpha and theta EEG bands (log scale). Different DL models converged to identify a similar set of 5-10 'preferred' features (>0.5 normalized activations), providing information about signal shape and logarithmic power distribution across frequency bands.

## 4. Discussion

We collected, preprocessed and analysed EEG from 11 participants across 4 stimulation and 1 control conditions. We found that delivering tDCS pulses in posterior cortical areas provided the most classifiable brain responses, with an F1-score of 92% on Holdout data. Comparatively, Holdout performance only reached 62% and 41% F1-scores for tACS and resting state conditions.

### 4.1 tACS vs. tDCS Effects

It is possible that effects induced by tACS may be more transient for relatively short durations[11,29], such as the 20-second blocks administered in our study. This would explain why our DL algorithms struggled to distinguish between resting state and tACS stimulation. Future studies could investigate the effects of varying stimulation/rest block durations from 5 to 30 seconds.

### 4.2 Feature Analysis

The top-5 most important features were common summary statistics—minimum, mode, and median values—as well as decibel power in alpha and theta EEG bands. In contrast, electrode importance analysis was non-conclusive due to varying spatial distributions. It is likely that DL algorithms learned intricate internal structure not readily available for human interpretation. Future work could find meaningful spatial patterns using substantially larger participant numbers or comparison with traditional ML algorithms.

### 4.3 Recommendations

Based on our findings, we advise future studies to choose tDCS stimulation over tACS (perhaps investigating additional types such as tRNS or tPCS). Regarding cortical regions, posterior areas resulted in slightly higher (+3%) classification accuracies, consistent with our initial hypothesis regarding brain controllability[18] and the "posterior cortical hot zone" identified as a reliable NCC.

### 4.4 Limitations

Given our relatively small final dataset (2.8 hours of aggregated eligible EEG), the full extent of validity and applicability of our results remains an open question. While outcomes reveal a promising research direction, clinical translation would require replicating results in a much larger dataset—at least an order of magnitude in terms of participants and hours of valid EEG data.

Based on the comprehensive systematic review by Roy et al.[1], we estimate that a robust dataset size for clinically viable performance would exceed 100 hours of valid training EEG data. A larger dataset would enable applying regularization techniques and help DL models learn the 'true signal' corresponding exclusively to TES-evoked brain responses.

## 5. Conclusions

This study demonstrated the feasibility of a framework using Deep Learning algorithms to classify EEG brain responses evoked by a defined pattern of TES neurostimulation. Our best CNN models reached 68.1% Holdout classification accuracy for all stimulation/control conditions—comparable to estimated human-level performance at 60-70%, based on clinical misdiagnosis rates[30]. Further research with larger EEG-TES datasets will be required to determine whether general Holdout performance can approach Training and Validation accuracies of >90%.

More importantly, we found that delivering tDCS pulses in posterior cortical areas—located inside the consciousness 'hot zone' identified in numerous recent publications[3,12,13]—elicited an exceptionally reliable brain response. Our CNN models achieved for this particular condition a remarkable **92% F1-score** (87% precision; 97% recall) on Holdout data from participants never seen during training, allowing us to confidently accept the central hypothesis: Deep Learning algorithms *can* distinguish brain responses evoked from different types of electrical stimulation in distant cortical regions.

#### Data and Code Availability

In the spirit of contributing to the progress of DL-EEG research, this project is open-sourced in full. Code is available on GitHub (github.com/alexispomares/DL-EEG-TES), and both raw and preprocessed EEG datasets can be found on Kaggle (kaggle.com/alexispomares/dissertation-raw and kaggle.com/alexispomares/dissertation-preprocessed).

## Acknowledgements

This work was supported by the Centre of Neurotechnology at Imperial College London. It builds on research conducted by Alexis Pomares Pastor under the supervision of Ines Ribeiro Violante and Gregory Scott.

The author declares no competing interests.